\documentclass[12pt]{iopart}
\usepackage{graphicx}
\begin{document}

\title{Thermoelectric properties of LaRh$_{1-x}$Ni$_x$O$_3$}

\author{S Shibasaki, Y Takahashi and I Terasaki}

\address{Department of Applied Physics, Waseda University, 
Tokyo 169-8555, Japan}
\ead{g01k0374@suou.waseda.jp}

\begin{abstract}
We report measurements and analyses of resistivity, thermopower, 
and thermal conductivity of polycrystalline samples of perovskite LaRh$_{1-x}$Ni$_x$O$_3$.
The thermopower is found to be large at 800~K (185 $\mu$V/K for 
$x=$0.3), which is ascribed to the high-temperature 
stability of the low-spin state of Rh$^{3+}$/Rh$^{4+}$ ions.
This clearly contrasts with the thermopower of the isostructural oxide LaCoO$_3$,
which rapidly decreases above 500~K owing to the spin-state transition.
The spin state of the transition-metal ions is one of the most important parameters in 
oxide thermoelectrics.

\end{abstract}

\pacs{72.20.Pa}
\submitto{\JPCM}
\maketitle

\section{Introduction}
One of the best advantages of oxides as thermoelectric materials is stability at high temperatures.
Thermoelectric materials are characterized by the dimensionless figure-of-merit, 
$ZT=S^2T/\rho\kappa$, where $S$, $\rho$, $\kappa$ and $T$ 
represent thermopower, resistivity, thermal conductivity and absolute temperature, respectively.
Na$_x$CoO$_2$\cite{NCO} was found to be useful for high-temperature power generation, 
and a large number of researchers now study thermoelectric oxides\cite{CCO1,CCO2,BSCO,LCO_thermal,LSCO,LCO,LnCO}.
In particular, the layered Co oxides with the CdI$_2$-type CoO$_2$ block 
show large thermopower up to 1000~K with low resistivity\cite{CCO1,CCO2,BSCO}.
Koshibae \etal successfully revealed that the spin and orbital degeneracy 
of the $d$ orbitals is important to the thermopower\cite{koshibae}.
According to their theoretical prediction, conduction between $(t_{2g})^5$ and $(t_{2g})^6$ causes large thermopower at high temperatures in Co oxides.
Rh oxides have been investigated for oxide thermoelectrics. Because Rh is just below Co in the periodic table, chemical properties of Rh are expected to be similar to those of Co.
We should note that Rh ions favour the low-spin state more than Co ions.
In fact, the layered Rh oxides with the CdI$_2$-type RhO$_2$ block are 
found to show similar thermoelectric properties to those of the layered Co oxides\cite{BSRO,BBRO,SRO,CRO_kuri,CRO}.

Although Rh is promising for oxide thermoelectrics, almost all the layered Rh oxides show similar but poorer thermoelectric properties than the Co oxides.
Here we focus on the perovskite-type Rh oxide LaRhO$_3$, because LaCo$_{1-x}$Ni$_x$O$_3$ and La$_{1-x}$Sr$_x$CoO$_3$ show good thermoelectric properties 
around room temperature\cite{LCO_thermal,LSCO,LCO,LnCO}.
However, the thermopower of LaCoO$_3$ suddenly decreases with the spin-state transition around 500~K\cite{LCO,LCO_to}, 
which means that LaCoO$_3$-based materials cannot be used above 500~K.
In contrast, Rh ions favour the low-spin state up to high temperature, and we expect that LaRhO$_3$-based
materials can show better thermoelectric properties than the Co analogues at high temperature.
We found that Rh-site substitution improves the thermoelectric properties in CuRhO$_{2}$\cite{CRO} and ZnRh$_{2}$O$_{4}$\cite{ZRO}.
Especially in the case of CuRhO$_2$, Mg substitution for Rh makes the system metallic, whereas Cu-site substitution does not increase conductivity.
Thus, we substituted Ni for Rh in LaRhO$_3$ as a reference material to LaCo$_{1-x}$Ni$_x$O$_3$.

In this paper, we present the high-temperature thermoelectric properties 
of perovskite LaRh$_{1-x}$Ni$_x$O$_3$ and compare them with those of LaCo$_{1-x}$Ni$_x$O$_3$.
We find that the thermopower of LaRhO$_3$ remains large above 500~K, and that
the spin state of the Co$^{3+}$/Rh$^{3+}$ ions determines the high-temperature thermoelectrics.

\section{Experimental details}
Polycrystalline samples of LaRh$_{1-x}$Ni$_x$O$_3$ were prepared by a solid-state reaction.
Stoichiometric amounts of La$_2$O$_3$, Rh$_2$O$_3$ and NiO were mixed and calcined at 1273~K for 24~h in air.
The calcined products were thoroughly ground, pelletized and sintered at 1373~K for 48~h in air.
The X-ray diffraction (XRD) of the samples was measured using a CuK$\alpha$ radiation by a $\theta$-$2\theta$ method from 10 to 100 degree.
The magnetic susceptibility measurements were performed using Magnetic Property Measurement System (MPMS, Quantum Design) with an external field of 0.1~T from 5 to 400 K.
The low-temperature resistivity and the thermopower measurements were performed in a liquid He cryostat from 4.2 to 300~K.
The high-temperature resistivity and thermopower measurements were performed in vacuum from 300 to 800~K.
The resistivity was measured using a conventional four-probe technique, and the thermopower was measured using a steady-state technique with a typical temperature gradient of 0.5 $-$ 1~K.
The thermal conductivity measurements were performed in a closed refrigerator using a steady-state technique from 8 to 300~K.

\section{Results and discussion}
\begin{figure}[tb]
\begin{center}
\includegraphics[width=8cm,clip]{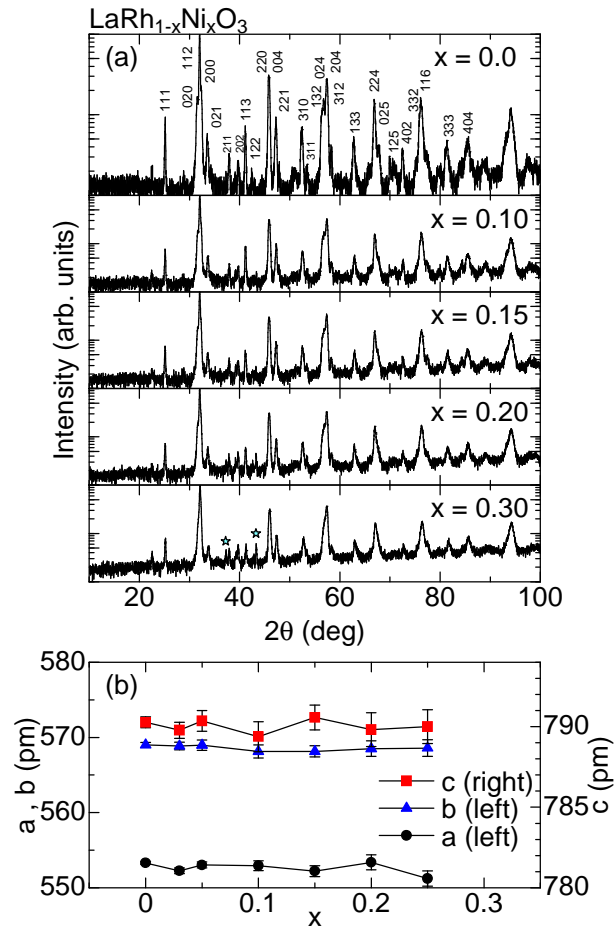}
\end{center}
\caption{\label{fig1}(Colour online)(a) XRD patterns and (b) lattice constants of LaRh$_{1-x}$Ni$_x$O$_3$. 
The impurity peaks of NiO are marked with stars.}
\end{figure}

\Fref{fig1}(a) shows the XRD patterns of LaRh$_{1-x}$Ni$_x$O$_3$ from $x=0$ to 0.3.
The lattice constants of these materials are shown in \fref{fig1}(b).
The samples are in single phase for $x\le$ 0.15, and a small amount of NiO is detected above $x=0.2$.
We evaluate the volume fraction of NiO impurity from the Rietveld simulation\cite{rietan} to be less than 4\% for $x=0.3$.
This suggests that more than 85\% of the doped Ni ions are substituted for Rh ions in the $x=0.3$ sample,
and we can safely neglect the effect of the NiO phase.
We further note that the resistivity and the thermopower systematically change 
up to $x=0.3$ (see \fref{fig3} and \fref{fig4}), suggesting that the NiO impurity 
little affects the thermoelectric properties.

\begin{figure}[tb]
\begin{center}
\includegraphics[width=7cm,clip]{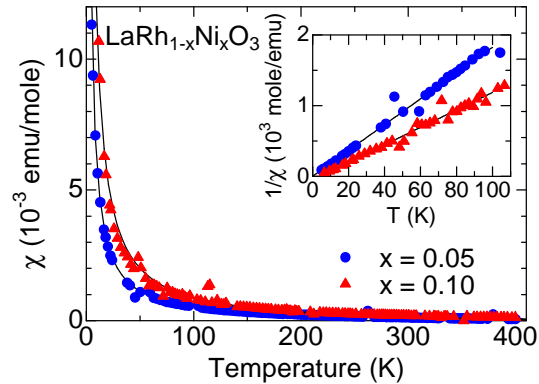}
\end{center}
\caption{\label{fig2}(Colour online) Susceptibility of LaRh$_{1-x}$Ni$_x$O$_3$. An external field of 0.1~T is applied. The solid lines are guides to the eye.}
\end{figure}

\begin{figure}[tb]
\begin{center}
\includegraphics[width=7cm,clip]{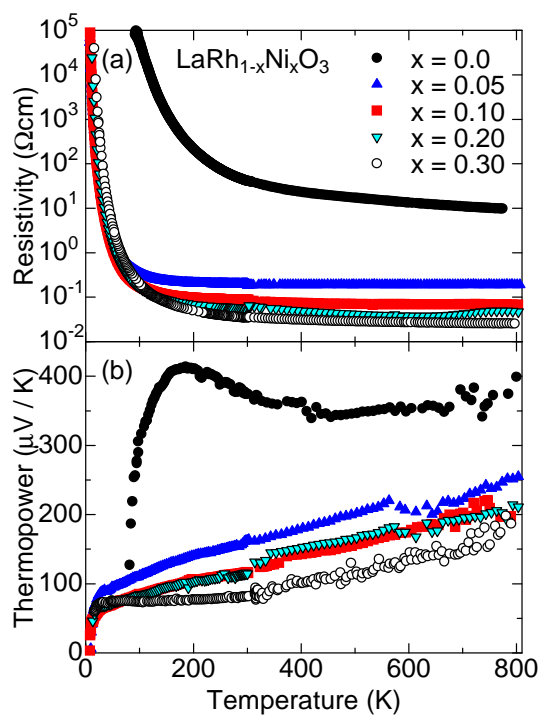}
\end{center}
\caption{\label{fig3}(Colour online)(a) Resistivity and (b) thermopower of LaRh$_{1-x}$Ni$_x$O$_3$ from 4.2 to 800~K.}
\end{figure}

\begin{figure}[tb]
\begin{center}
\includegraphics[width=7cm,clip]{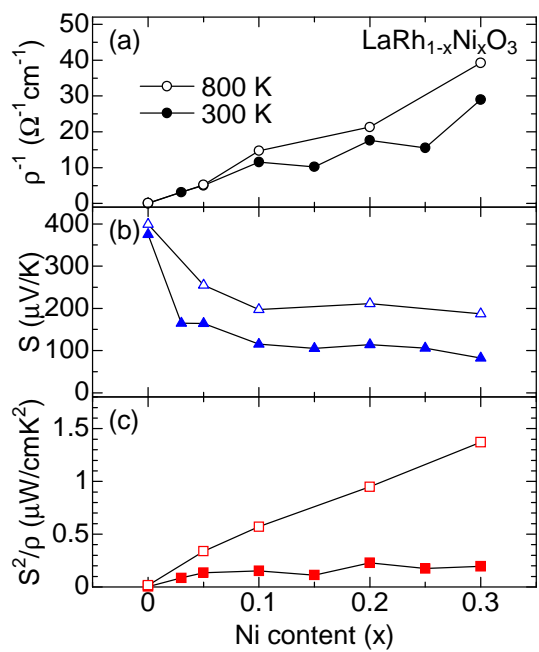}
\end{center}
\caption{\label{fig4}(Colour online)Ni content ($x$) dependence of 
electric conductivity ($\rho^{-1}$), 
thermopower ($S$) and power factor ($S^2/\rho$)
of LaRh$_{1-x}$Ni$_x$O$_3$ at 300~K (filled) and 800~K (unfilled). }
\end{figure}

In our sintering condition, we expect that Ni ions are stable as divalent.
On the other hand, a possible existence of Ni$^{3+}$ was reported in La$_2$NiRhO$_6$\cite{LNRO}, 
and hence we measured susceptibility of LaRh$_{1-x}$Ni$_x$O$_3$ in order to determine the valence state of Ni.
\Fref{fig2} shows the susceptibility of LaRh$_{1-x}$Ni$_x$O$_3$ from 5 to 400~K.
The inset of \fref{fig2} shows the inverse susceptibility of LaRh$_{1-x}$Ni$_x$O$_3$.
We obtain Curie constants of 6.10$\times 10^{-2}$ and 1.11$\times 10^{-1}$~emu/mole$\cdot$K for $x=0.05$ and $0.1$ using the Curie law, respectively.
Adopting the $g$-factor of 2, we calculate Curie constants assuming Ni$^{2+}$ in the high-spin state to be 5.0$\times 10^{-2}$ and 1.0$\times 10^{-1}$~emu/mole$\cdot$K for $x=0.05$ and $x=0.1$, respectively.
From these values, we conclude that substituted Ni ions are divalent (high spin) in our system.
Conduction occurs in the Rh-O network, and it is hard to see the magnetic susceptibility of Rh ions.
Nakamura \etal showed that both Rh$^{3+}$ and Rh$^{4+}$ ions are in the low-spin state\cite{LRO}.
Assuming that the oxygen content does not change largely, we expect that 
the substitution of Ni$^{2+}$ (the ion radius \cite{shannon} $r=$0.70~\AA) 
for Rh$^{3+}$ ($r=$0.67~\AA) creates Rh$^{4+}$ ($r=$0.62~\AA) owing to 
the charge neutrality condition.
In other words, two Rh$^{3+}$ ions are replaced by Ni$^{2+}$ and Rh$^{4+}$ ions through this substitution.
Then the ``average'' ion radius of the dopant is 
(0.70+0.62)/2=0.66 \AA, which is nearly equal to the ion radius
of Rh$^{3+}$.
Accordingly the lattice parameters are expected to depend weakly on $x$,
 which is consistent with the data in \fref{fig1}(b).
A similar tendency is observed in CuRh$_{1-y}$Mg$_y$O$_2$, 
where Mg$^{2+}$(0.72~\AA) ions are substituted for Rh$^{3+}$ 
to create Rh$^{4+}$\cite{CRO}.

Now let us discuss the thermoelectric properties of LaRh$_{1-x}$Ni$_x$O$_3$.
\Fref{fig3}(a) shows the resistivity of LaRh$_{1-x}$Ni$_x$O$_3$ from 4.2 to 800~K.
Nakamura \etal previously reported that the resistivity of LaRhO$_3$ 
was 45~$\Omega$cm at room temperature with semiconducting behaviour\cite{LRO}, 
which is reproduced in our measurement.
The resistivity systematically decreases with $x$ and reaches as low as 25~m$\Omega$cm at 800~K for $x=0.3$.
The insulating behaviour of the resistivity at low temperature suggests that the substituted Ni ions work as scattering centers.
This further assures us to regard the Ni ions as divalent.
If the Ni ions were trivalent, the system could be regarded as a solid solution of LaRhO$_3$ and LaNiO$_3$, in which the resistivity would be dominated by the volume fraction of metallic LaNiO$_3$.
\Fref{fig3}(b) shows the thermopower of LaRh$_{1-x}$Ni$_x$O$_3$ from 4.2 to 800~K.
After the thermopower drastically deareases from $x=0$ to $0.05$, it gradually decreases with $x$ above $x=0.1$.
We should note that the sign is always positive for all the samples, which further excludes the possible existence of Ni$^{3+}$ ions giving negative thermopower\cite{LNO}.
We observe large thermopower up to 800~K, which suggests that Rh ions are in the low-spin state up to high temperature.

\Fref{fig4} shows the electric conductivity ($\rho^{-1}$), 
the thermopower ($S$) and the power factor ($S^2/\rho$) at 300 and 800~K as
a function of Ni content $x$. 
We clearly see that the conductivity increases with $x$, which suggests that Ni is a suitable dopant.
The thermopower is weakly dependent on $x$ at high concentration.
Thanks to these behaviours, we obtain fairly large power factor which increases up to $x=0.3$ in LaRh$_{1-x}$Ni$_x$O$_3$.
Recently, Usui \etal theoretically calculated the thermopower of doped LaRhO$_{3}$ as a function of carrier concentration\cite{Usui}, which agrees well with our experiment.
This is quantitatively different from the power factor of conventional semiconductors, 
where it takes a maximum at an optimum carrier density of 10$^{19}$-10$^{20}$ cm$^{-3}$.
The thermoelectric parameters are listed for LaRh$_{0.7}$Ni$_{0.3}$O$_3$ and LaCo$_{0.8}$Ni$_{0.2}$O$_3$ in \tref{tab1}.
While the resistivity of LaRh$_{0.7}$Ni$_{0.3}$O$_3$ is 25 times higher than that of LaCo$_{0.8}$Ni$_{0.2}$O$_3$, 
its thermopower is 12 times larger at 800~K.
This indicates that LaRh$_{0.7}$Ni$_{0.3}$O$_3$ shows a larger power factor at 800~K.

\begin{table}
 \caption{\label{tab1}Comparison between LaRh$_{0.7}$Ni$_{0.3}$O$_3$ 
 and LaCo$_{0.8}$Ni$_{0.2}$O$_3$.
 The subscripts represent measured temperatures.}
 \begin{indented}
 \lineup
 \item[]\begin{tabular}{@{}lrr}
 \br
  &  LaRh$_{0.7}$Ni$_{0.3}$O$_3$ & LaCo$_{0.8}$Ni$_{0.2}$O$_3$ \\
 \mr
 $\rho_{\mbox{\scriptsize{ 300~K}}}$ (m$\Omega$cm) & \036 \qquad & \010$^{\rm a}$ \qquad \cr
 $\rho_{\mbox{\scriptsize{ 800~K}}}$ (m$\Omega$cm) & \025 \qquad & \0\01$^{\rm b}$ \qquad \cr
 $S_{\mbox{\scriptsize{ 300~K}}}$ ($\mu$V/K) & \085 \qquad & 100$^{\rm a}$ \qquad \cr
 $S_{\mbox{\scriptsize{ 800~K}}}$ ($\mu$V/K) & 185 \qquad & \015$^{\rm b}$ \qquad \cr
 $S^2/\rho_{\mbox{\scriptsize{ 800~K}}}$ ($\mu$W/cmK$^2$) & 1.37 \qquad & 0.23$^{\rm b}$ \qquad \cr
 $\kappa_{\mbox{\scriptsize{ 300~K}}}$ (mW/cmK) & \020 \qquad & \014$^{\rm a}$ \qquad \cr
 \br
 \end{tabular}
 \item[] $^{\rm a}$ Reference \cite{LCO_thermal}.
 \item[] $^{\rm b}$ Reference \cite{LCO}.
 \end{indented}
\end{table}

\begin{figure}[tb]
\begin{center}
\includegraphics[width=7cm,clip]{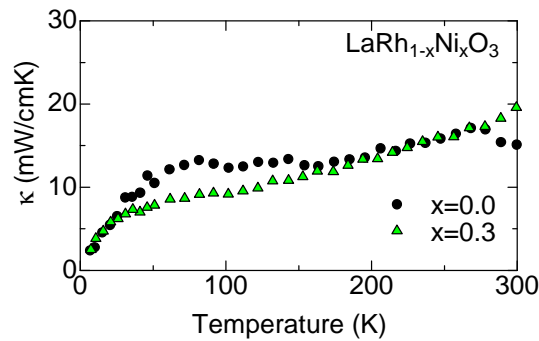}
\end{center}
\caption{\label{fig5}(Colour online) Thermal conductivity of LaRh$_{1-x}$Ni$_x$O$_3$ from 8 to 300~K.}
\end{figure}

The thermal conductivity is essential to evaluating $ZT$.
\Fref{fig5}(a) shows the thermal conductivity of LaRh$_{1-x}$Ni$_x$O$_3$ below 300~K.
The magnitude of the thermal conductivity of LaRhO$_3$ and LaRh$_{0.7}$Ni$_{0.3}$O$_3$ 
is almost the same as that of LaCo$_{1-x}$Ni$_x$O$_3$\cite{LCO_thermal}.
The electronic contribution is evaluated to be less than 1~mW/cmK at 300~K 
for LaRh$_{0.7}$Ni$_{0.3}$O$_3$ from the Wiedemann-Franz law, which means that the lattice contribution is dominant.
$ZT$ is calculated to be 3$\times$10$^{-3}$ for $x=0.3$ 
and 6$\times$10$^{-5}$ for $x=0$ at 300~K.
The lattice thermal conductivity is dominant in both materials and the magnitude is almost same, 
we assume the thermal conductivity of LaCoO$_3$ (25~mW/cmK at 800~K)\cite{LnCO} and evaluate $ZT$ to be 0.044 for $x=0.3$.
We notice that this value itself is not yet satisfactory, but should emphasize that it is three times larger than 
$ZT=$0.015 for LaCo$_{0.95}$Ni$_{0.05}$O$_3$ at 800~K\cite{LnCO}.
As far as we know, this is the first report for a Rh oxide to show better thermoelectric properties than the isostructural Co oxide.
Assuming that the lattice thermal conductivity is zero, we find $ZT=S^2T/\kappa_{\rm el}\rho=S^2/L_0$ by using the Wiedemann-Franz law, where $L_0$ is the Lorentz number\cite{ZT_singh}.
Then $ZT>1$ requires $S>160$~$\mu$V/K, which is satisfied in the present compound at 800~K.
The thermopowers of the layered Co/Rh oxides and LaRh$_{1-x}$Ni$_x$O$_3$ exceed this value at high temperature.
This comes from the stability of the low-spin state of the Co/Rh ions\cite{koshibae}.
On the other hand, LaCoO$_3$-based materials show the thermopower less than 30~$\mu$V/K at high temperature, which means that these compounds do not show $ZT>1$ at high temperature.
Thus we conclude that the spin state of Co/Rh ions in Co/Rh oxides plays a crucial role in oxide thermoelectrics.
Unfortunately, the resistivity of LaRh$_{0.7}$Ni$_{0.3}$O$_3$ is still high, and hence $ZT$ is low.
This is perhaps because we substituted the Rh site, not the La site.
We expect better thermoelectric properties in La-site substituted LaRhO$_3$.

\section{Summary}
In summary, we have presented the transport data of LaRh$_{1-x}$Ni$_x$O$_3$, 
and have compared them with those of LaCo$_{1-x}$Ni$_x$O$_3$.
Unlike LaCo$_{1-x}$Ni$_x$O$_3$, the thermopower remains large up to 800~K, 
which is ascribed to the conduction between the low-spin states of Rh$^{3+}$ and Rh$^{4+}$ ions.
The dimensionless figure-of-merit is evaluated to be 0.044 for $x=0.3$ 
at 800~K, which is almost three times larger than that for Ni-doped LaCoO$_3$.
We propose that the spin-state control is a unique strategy for
thermoelectric-materials design in transition metal oxides.

\ack
The authors would like to thank T Nakano and Y Klein for fruitful discussion.
They would also like to thank S Yoshida for technical support.
This work was partially supported by a Grant-in-Aid for JSPS Fellows.

\end{document}